\documentstyle[twoside,fleqn,npb,epsfig,axodraw]{article}
\newcommand{\AmS}{{\protect\the\textfont2
  A\kern-.1667em\lower.5ex\hbox{M}\kern-.125emS}}

\newcommand{\lsim}
{\mathrel{\raisebox{-.3em}{$\stackrel{\displaystyle <}{\sim}$}}}

\def\asymp#1%
{\mathrel{\raisebox{-.4em}{$\widetilde{\scriptstyle #1}$}}}

\def\Nequal#1%
{\mathrel{\raisebox{-.5em}{$\stackrel{=}{\scriptstyle\rm#1}$}}}
\newcommand{\dsl}[1]{\not \hspace{-0.7mm}#1}
\def\dsl{\mathpalette\make@slash}
\def\make@slash#1#2{\setbox\z@\hbox{$#1#2$}%
  \hbox to 0pt{\hss$#1/$\hss\kern-\wd0}\box0}

\def\beq{\begin{equation}}
\def\eeq{\end{equation}}
\def\beqar{\begin{eqnarray}}
\def\eeqar{\end{eqnarray}}
\def\barr#1{\begin{array}{#1}}
\def\earr{\end{array}}
\def\bfi{\begin{figure}}
\def\efi{\end{figure}}
\def\btab{\begin{table}}
\def\etab{\end{table}}
\def\bce{\begin{center}}
\def\ece{\end{center}}
\def\nn{\nonumber}

\def\text{\textstyle}

\def\al{\alpha}

\def\ga{\gamma}
\def\de{\delta}

\def\si{\sigma}

\def\reffi#1{\mbox{Fig.~\ref{#1}}}
\def\reffis#1{\mbox{Figs.~\ref{#1}}}
\def\refta#1{\mbox{Table~\ref{#1}}}

\def\citere#1{\mbox{Ref.~\cite{#1}}}
\def\citeres#1{\mbox{Refs.~\cite{#1}}}

\newcommand{\GeV}{\unskip\,\mathrm{GeV}}
\newcommand{\MeV}{\unskip\,\mathrm{MeV}}

\newcommand{\fb}{\unskip\,\mathrm{fb}}

\newcommand{\ri}{{\mathrm{i}}}
\newcommand{\rd}{{\mathrm{d}}}

\newcommand{\M}{{\cal{M}}}
\def\mathswitchr#1{\relax\ifmmode{\mathrm{#1}}\else$\mathrm{#1}$\fi}

\newcommand{\PW}{\mathswitchr W}

\newcommand{\Pe}{\mathswitchr e}

\newcommand{\Pd}{\mathswitchr d}

\newcommand{\Pu}{\mathswitchr u}
\newcommand{\Pubar}{\bar{\mathswitchr u}}

\newcommand{\Pep}{\mathswitchr {e^+}}
\newcommand{\Pem}{\mathswitchr {e^-}}
\newcommand{\PWp}{\mathswitchr {W^+}}
\newcommand{\PWm}{\mathswitchr {W^-}}

\def\mathswitch#1{\relax\ifmmode#1\else$#1$\fi}

\newcommand{\MW}{\mathswitch {M_\PW}}

\newcommand{\GW}{\Gamma_{\PW}}

\newcommand{\GF}{\mathswitch {G_\mu}}

\def\solid{\raise.9mm\hbox{\protect\rule{1.1cm}{.2mm}}}
\def\dash{\raise.9mm\hbox{\protect\rule{2mm}{.2mm}}\hspace*{1mm}}

\def\ie{i.e.\ }

\newcommand{\eeWW}{{\Pe^+ \Pe^-\to \PW^+ \PW^-}}

\newcommand{\eeWWffff}{\Pep\Pem\to\PW\PW\to 4f}

\newcommand{\eeffff}{\Pep\Pem\to 4f}
\newcommand{\eeffffg}{\eeffff\ga}
\renewcommand{\O}{{\cal O}}

\def\draftdate{\relax}
\def\mda{\relax}
\def\mua{\relax}
\def\mla{\relax}
\def\draft{
\def\thtystars{******************************}
\def\sixtystars{\thtystars\thtystars}
\typeout{}
\typeout{\sixtystars**}
\typeout{* Draft mode!
         For final version remove \protect\draft\space in source file *}
\typeout{\sixtystars**}
\typeout{}
\def\draftdate{\today}
\def\mua{\marginpar%[\boldmath\hfil$\uparrow$]%
                   {\boldmath$\uparrow$\hfil}%
                    \typeout{marginpar: $\uparrow$}\ignorespaces}
\def\mda{\marginpar%[\boldmath\hfil$\downarrow$]%
                   {\boldmath$\downarrow$\hfil}%
                    \typeout{marginpar: $\downarrow$}\ignorespaces}
\def\mla{\marginpar%[\boldmath\hfil$\rightarrow$]%
                   {\boldmath$\leftarrow $\hfil}%
                    \typeout{marginpar: $\leftrightarrow$}\ignorespaces}
\def\Mua{\marginpar%[\boldmath\hfil$\Uparrow$]%
                   {\boldmath$\Uparrow$\hfil}%
                    \typeout{marginpar: $\uparrow$}\ignorespaces}
\def\Mda{\marginpar%[\boldmath\hfil$\Downarrow$]%
                   {\boldmath$\Downarrow$\hfil}%
                    \typeout{marginpar: $\downarrow$}\ignorespaces}
\def\Mla{\marginpar%[\boldmath\hfil$\Rightarrow$]%
                   {\boldmath$\Leftarrow $\hfil}%
                    \typeout{marginpar: $\leftrightarrow$}\ignorespaces}
\overfullrule 5pt
\oddsidemargin -15mm
\marginparwidth 29mm
}

\def\stars{\strut\leaders\hbox{*}\hfill\strut}
\def\starline{\hfil\strut\hfil\hbox to \textwidth {\stars}\hfil}

\title{Radiative corrections to $\eeWWffff$ with {\sc RacoonWW}}

\author{  
  A.~Denner\address{Paul-Scherrer-Institut, Villigen, Switzerland},
  S.~Dittmaier\address{Universit\"at Bielefeld, Bielefeld, Germany}%
    \thanks{Partially supported by the Bundesministerium f\"ur Bildung und
    Forschung, No.~05~7BI92P~9, Bonn, Germany.},
  M.~Roth\address{Universit\"at Leipzig, Leipzig, Germany} and 
  D.~Wackeroth\address{University of Rochester, Rochester NY, USA}
}

\begin{document}

%%%

\thispagestyle{empty}
\def\thefootnote{\fnsymbol{footnote}}
\setcounter{footnote}{1}
\hfill BI-TP 2000/15 \\
\vspace*{3cm}
\begin{center}
{\Large\bf 
Radiative corrections to $\eeWWffff$ with {\sc RacoonWW}%
\footnote{To appear in the 
{\it Proceedings of the 5th Zeuthen Workshop on Elementary
Particle Theory ``Loops and Legs in Quantum Field Theory''},
Bastei/K\"onigstein, Germany, April 9-14, 2000.}
}
\\[1.8cm]
{\large
A.~Denner$^1$,
S.~Dittmaier$^2$%
\footnote{Partially supported by the Bundesministerium f\"ur Bildung und
Forschung, No.~05~7BI92P~9, Bonn, Germany.},
M.~Roth$^3$ and D.~Wackeroth$^4$}\\[1.5em]
\parbox{8cm}{$^1$Paul-Scherrer-Institut, Villigen, Switzerland\\[.5em]
$^2$Universit\"at Bielefeld, Bielefeld, Germany\\[.5em]
$^3$Universit\"at Leipzig, Leipzig, Germany\\[.5em]
$^4$University of Rochester, Rochester NY, USA}
\end{center}
\vspace*{3.0cm}
{\large\bf Abstract}\\[.2cm]
\setlength{\baselineskip}{8pt}
{\sc RacoonWW} is the first Monte Carlo generator for
$\eeWWffff(+\gamma)$ that includes the electroweak ${\cal O}(\alpha)$ 
radiative corrections in the double-pole approximation completely.
Some numerical results for LEP2 energies are discussed, and the
predictions for the total W-pair cross section are confronted with LEP2
data.
\par
\vskip 4cm
\noindent
May 2000        
\null

\clearpage

\def\thefootnote{\fnsymbol{footnote}}
\setcounter{footnote}{1}
\setcounter{page}{1}
%%%

\begin{abstract}
{\sc RacoonWW} is the first Monte Carlo generator for
$\eeWWffff(+\gamma)$ that includes the electroweak ${\cal O}(\alpha)$ 
radiative corrections in the double-pole approximation completely.
Some numerical results for LEP2 energies are discussed, and the
predictions for the total W-pair cross section are confronted with LEP2
data.
\end{abstract}

% typeset front matter (including abstract)
\maketitle

\section{W-PAIR PRODUCTION AT LEP2}

The investigation of W-pair production at LEP2 plays an important role
in the verification of the Electroweak Standard Model (SM). 
Apart from the direct observation of the triple-gauge-boson
couplings in $\eeWW$, the increasing accuracy in the W-pair-production
cross-section and W-mass measurements 
has put this process into the row of SM precision tests. 

To account for the high experimental accuracy \cite{lep2res} on the
theoretical side is a great challenge: the W bosons have to be treated
as resonances in the full four-fermion processes $\eeffff$,
and radiative corrections need to be included. While 
several lowest-order
predictions are based on the full set of Feynman diagrams, 
only very few calculations include radiative
corrections beyond the level of universal effects (see 
\citeres{lep2repWcs,lep2mcws} and references therein).
Fortunately, to match the experimental precision for W-pair production at LEP2 
a full one-loop calculation for the four-fermion processes is not needed,
and it is sufficient to take into account only those radiative corrections 
that are enhanced by two resonant W bosons.  
The neglected corrections are of the order $(\al/\pi)(\GW/\MW)$, \ie
below 0.5\% even if possible enhancement factors are taken into account.
The theoretically clean way to carry out this approximation is
the expansion about the two resonance poles, which is called {\it
double-pole approximation} (DPA) \cite{st91}.
A full description of this strategy and of different variants used in
the literature \cite{yfsww,Be98,ku99}
(some of them involving further approximations) is beyond the scope of
this article. We can only briefly sketch the approach pursued in
{\sc RacoonWW} \cite{racoonww_lep2res,De00}.

\section{RADIATIVE CORRECTIONS WITH {\sc RacoonWW}}

In DPA, ${\cal O}(\alpha)$ corrections to
$\Pep\Pem\to\PW\PW\to 4f$ can be classified into two types:
factorizable and non-factorizable corrections.
We first focus on virtual corrections.

{\it Factorizable corrections} are those that correspond either 
to W-pair production or to W~decay. Virtual factorizable corrections 
are represented by the schematic diagram of 
\reffi{fig:fdiag}, in which the shaded blobs contain all one-loop
corrections to the on-shell production and on-shell 
decay processes, and the open blobs
include the corrections to the W~propagators. 
\begin{figure}
{\centerline{
\unitlength 1pt
\begin{picture}(190,90)(0,10)
\ArrowLine(30,50)( 5, 95)
\ArrowLine( 5, 5)(30, 50)
\Photon(30,50)(150,80){2}{11}
\Photon(30,50)(150,20){2}{11}
\ArrowLine(150,80)(190, 95)
\ArrowLine(190,65)(150,80)
\ArrowLine(190, 5)(150,20)
\ArrowLine(150,20)(190,35)
\GCirc(30,50){10}{.5}
\GCirc(90,65){10}{1}
\GCirc(90,35){10}{1}
\GCirc(150,80){10}{.5}
\GCirc(150,20){10}{.5}
\DashLine( 70,10)( 70,90){2}
\DashLine(110,10)(110,90){2}
\put(40,26){$W$}
\put(40,63){$W$}
\put(120, 6){$W$}
\put(120,83){$W$}
\end{picture}
} }
\caption{Structure of virtual factorizable corrections with
loop corrections in the blobs}
\label{fig:fdiag}
\end{figure}
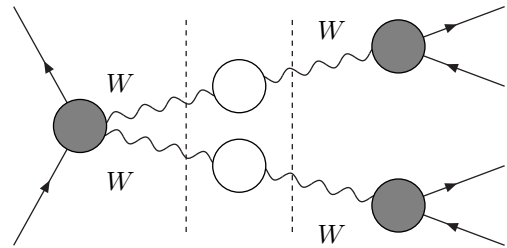
\begin{figure}
{\centerline{
\unitlength 1pt
\begin{picture}(120,90)(0,10)
\ArrowLine(30,50)( 5, 95)
\ArrowLine( 5, 5)(30, 50)
\Photon(30,50)(90,80){2}{6}
\Photon(30,50)(90,20){2}{6}
\GCirc(30,50){ 7}{0.5}
\Vertex(90,80){1.2}
\Vertex(90,20){1.2}
\ArrowLine(90,80)(120, 95)
\ArrowLine(120,65)(105,72.5)
\ArrowLine(105,72.5)(90,80)
\Vertex(105,72.5){1.2}
\ArrowLine(120, 5)( 90,20)
\ArrowLine( 90,20)(105,27.5)
\ArrowLine(105,27.5)(120,35)
\Vertex(105,27.5){1.2}
\Photon(105,27.5)(105,72.5){2}{4.5}
\put(93,47){$\gamma$}
\put(55,73){$W$}
\put(55,16){$W$}
\end{picture}
} }
\caption{ Typical diagram for virtual non-factorizable corrections}
\label{fig:nfdiag}
\end{figure}
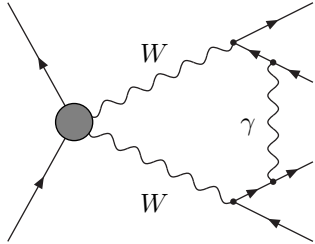
For the corresponding matrix element ${\cal M}$ the application of the
DPA amounts to the replacement
\newcommand{\kp}{k_{\PW^+}}
\newcommand{\km}{k_{\PW^-}}
\beqar
{\cal M} &=& \frac{R(\kp^2,\km^2)}{(\kp^2-\MW^2)(\km^2-\MW^2)} 
\nn\\
&\to&
%\frac{R(\MW^2,\MW^2)}{(\kp^2-\MW^2+\ri\MW\GW)(\km^2-\MW^2+\ri\MW\GW)},
\frac{R(\MW^2,\MW^2)}{(\kp^2-M^2)(\km^2-M^2)},
\eeqar
where the originally gauge-dependent numerator $R(\kp^2,\km^2)$ 
is replaced by the gauge-indepen\-dent residue $R(\MW^2,\MW^2)$, and
$M^2=\MW^2-\ri\MW\GW$ is the location of the poles in the complex 
$k_{\PW^\pm}^2$ planes.
The one-loop corrections to this residue can be deduced from the known
results for the pair production \cite{wwprod} 
and the decay \cite{wdecay} of on-shell W~bosons.
However, the spin correlations between the two W~decays should be taken
into account.

{\it Non-factorizable corrections} \cite{be97,de98}
comprise all those doubly-resonant
corrections that are not yet contained in the factorizable ones, 
and include, in particular, all diagrams involving particle exchange between the
subprocesses. Such diagrams only lead to
doubly-resonant contributions if the exchanged particle is a photon
with energy $E_\gamma\lsim\Gamma_\PW$; all other non-factorizable
diagrams are negligible in DPA. A typical diagram for a virtual 
non-factorizable correction is shown in
\reffi{fig:nfdiag}, where the full blob represents tree-level subgraphs.
We note that diagrams involving photon exchange between the W~bosons
contribute to both factorizable and
non-factorizable corrections; otherwise the splitting into those parts
would not be gauge-invariant.

In {\sc RacoonWW} the virtual corrections are treated in DPA, including
the full set of factorizable and non-factorizable ${\cal O}(\alpha)$ 
corrections. The real corrections are calculated from full 
matrix elements for $\eeffffg$, as described in \citere{ee4fa},
i.e.\ the DPA is not used in this part.
In this way, we avoid potential problems in the definition of the DPA
for the emission of photons with energies $E_\gamma\sim\Gamma_\PW$.
On the other hand, this asymmetry in the calculation of virtual and real
corrections requires particular care concerning the structure of IR and
mass singularities. The singularities have the form of a
universal radiator function convoluted with the
respective lowest-order matrix element squared $|\M_0|^2$ 
of the non-radiative process.  
Since the virtual corrections are calculated in DPA, but the full
matrix element is used for the real photons, a simple summation
of virtual and real corrections would lead to a
mismatch in the singularity structure and eventually to totally wrong
results. Therefore, we extract those singular
parts from the real photon contribution that exactly match the
singular parts of the virtual photon contribution, then replace in
these terms the full $|\M_0|^2$ 
by the one calculated in DPA, and finally add
this modified part to the virtual corrections. This modification is
allowed within DPA and leads to a proper
matching of all IR and mass singularities. This treatment has been
carried out in two different ways, once following phase-space
slicing, once using the subtraction formalism of \citere{subtract}.
\looseness -1

Beyond ${\cal O}(\alpha)$, {\sc RacoonWW} includes soft-photon
exponentiation and leading higher-order ISR effects in the
structure-function approach. 
Using $\GF$ as input parameter instead of $\alpha(0)$ in the 
lowest-order cross section, also the leading 
higher-order effects from $\Delta\alpha$ and
$\Delta\rho$ are taken into account. On the other hand, in the
relative correction factor we use $\alpha(0)$ in order to describe
the couplings of the real photons correctly.

\section{NUMERICAL RESULTS}

\subsection{Predictions from {\sc RacoonWW}}

A survey of numerical results obtained with {\sc RacoonWW} has already
been presented in \citere{racoonww_lep2res} for LEP2 and linear-collider
energies. Here we only review the W~invariant-mass distribution 
given there, extend the results for the total cross section, 
and add some studies of the intrinsic ambiguities in the DPA.

\begin{figure}%
{\centerline{ 
\setlength{\unitlength}{1cm}
\begin{picture}(7.7,7.3)
\put(-.4,-.6){\includegraphics{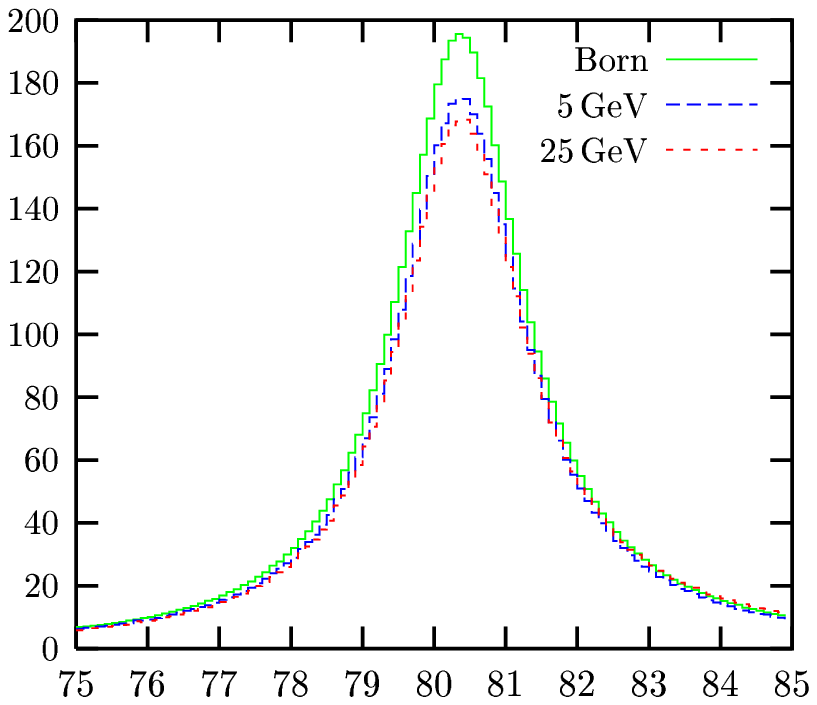}}
\put(0.1,6.4){\makebox(1,1)[l]{$\frac{\rd \si}{\rd M_{\Pd\Pu}}\
\left[\frac{\fb}\GeV\right]$}}
\put(4.0,-0.8){\makebox(1,1)[c]{$M_{\Pd\Pu}\ [\mathrm{GeV}]$}}
\end{picture}
}} 
\caption{Invariant-mass distribution of the $\Pd\Pubar$ pair for
  $\Pep\Pem\to\nu_\mu\mu^+\Pd\Pubar$ and $\protect\sqrt{s}=200\GeV$
(taken from \protect\citere{racoonww_lep2res}a)}
\label{fi:ud_invmass_abs}
\end{figure}

Figure~\ref{fi:ud_invmass_abs} 
shows the invariant-mass distribution
of the $\Pd\Pubar$ quark pair for the semi-leptonic channel
$\Pep\Pem\to\nu_\mu\mu^+\Pd\Pubar$ at $\sqrt{s}=200\GeV$ at 
tree-level and with electroweak $\O(\alpha)$ corrections for two different
recombination cuts, $M_{\mathrm{rec}}=5\GeV$ 
and $25\GeV$.  
\begin{figure}%
{\centerline{ 
\setlength{\unitlength}{1cm}
\begin{picture}(7.7,7.3)
\put(0.5,6.4){\makebox(1,1)[c]{$\de\ [\%]$}}
\put(4.0,-0.8){\makebox(1,1)[c]{$M_{\Pd\Pu}\ [\mathrm{GeV}]$}}
\put(-.4,-.6){\includegraphics{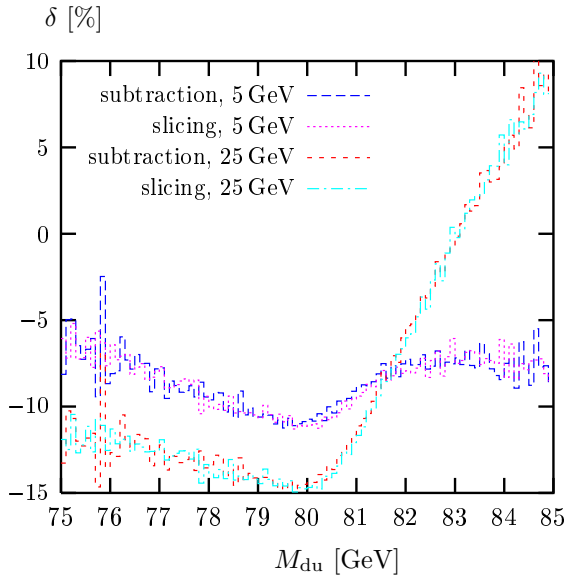}}
\end{picture}
} } 
\caption{Relative corrections to the invariant-mass distribution of the $\Pd\Pubar$ pair for
  $\Pep\Pem\to\nu_\mu\mu^+\Pd\Pubar$ and $\protect\sqrt{s}=200\GeV$
  (taken from \protect\citere{racoonww_lep2res}a)}
\label{fi:ud_invmass_rel}
\end{figure}%
The recombination of photons with final-state charged fermions is
performed as described in \citere{racoonww_lep2res}: we first determine
the lowest invariant mass $M_{\gamma f}$ built by the emitted photon
and a charged final-state fermion. If $M_{\gamma f}$ is smaller than 
$M_{\mathrm{rec}}$, the photon momentum is
added to the one of the corresponding fermion $f$.  
The locations of the maxima in the
corrected line shapes differ by up to \mbox{30}$\MeV$ between the two
values of $M_{\mathrm{rec}}$. As expected, there is a tendency to
shift the maxima to larger invariant masses if more and more photons
are recombined. In Fig.~\ref{fi:ud_invmass_rel} we display the
relative corrections $\delta=\rd\sigma/\rd\si_0-1$ for the two values of
$M_{\mathrm{rec}}$, which illustrates the strong dependence of the
corrected invariant-mass distributions on the treatment of the real
photons.  We obtain consistent results for the phase-space 
``slicing'' and the ``subtraction'' methods. The size of
the shown effects demonstrates that a careful treatment of real
photons is mandatory in the W-mass reconstruction at LEP2 accuracy.
\looseness -1

\begin{figure}
\setlength{\unitlength}{1cm}
{\centerline{
\begin{picture}(9,7.4)
\put(.1,-1.3){\includegraphics{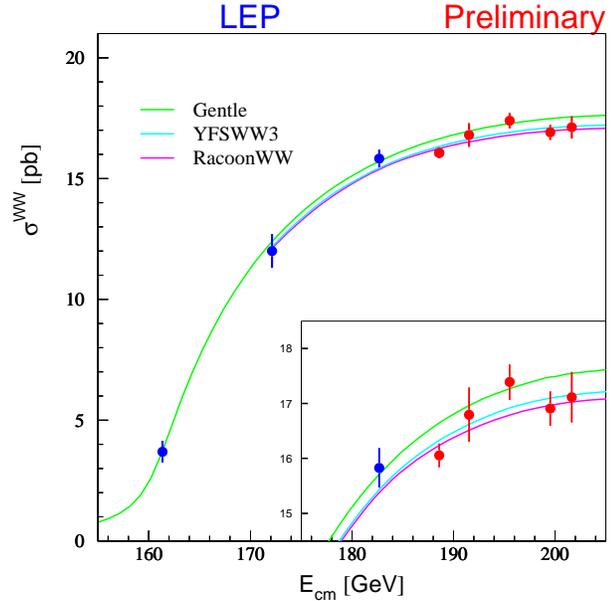}}
\end{picture}
} } 
\caption{Total WW production cross section at LEP2, as given by the
LEPEWWG \protect\cite{LEPEWWG}}
\label{fig:wwcs2000}
\end{figure}

Figure~\ref{fig:wwcs2000} shows a comparison of {\sc RacoonWW} results and 
of other predictions with recent LEP2 data, as given by the LEP Electroweak
Working Group \cite{lep2res,LEPEWWG}.
The data are in good agreement with the predictions of {\sc RacoonWW}
and {\sc YFSWW3} \cite{yfsww}.
The predictions of these two generators differ between 0.5--0.7\%.%
\footnote{Meanwhile the dominant source of this difference has been
  found, and the new results of YFSWW3 
%are closer to the results of
agree within 0.3\% with the results of
  {\sc RacoonWW}. Details on the new YFSWW3 predictions can be found
  in \citere{lep2mcws}.}
More details on the conceptual differences of the two generators, as
well as a detailed comparison of numerical results, can be found in
\citere{lep2mcws}.
Figure~\ref{fig:wwcs2000} also includes the prediction provided by
{\sc GENTLE} \cite{gentle}, which differs from the {\sc RacoonWW} and
{\sc YFSWW3} results by 2--2.5\%. This difference is due to the neglect
of non-leading, non-universal ${\cal O}(\alpha)$ corrections in
{\sc GENTLE}. 
Consequently, the comparison between SM predictions 
with the precise measurements of the W-pair production cross section
at LEP2 reveals evidence of non-leading electroweak radiative
corrections beyond the level of universal effects.

\subsection{Intrinsic ambiguities of the DPA}

In order to investigate the accuracy of the DPA quantitatively, we have 
performed a number of tests.
The implementation of the DPA has been modified
within the formal level of $\alpha\Gamma_\PW/\MW$,
and the obtained results have been compared.
Because in {\sc RacoonWW} only the virtual corrections are 
treated in DPA, while real photon emission is based on the full
$\eeffffg$ matrix element with the exact five-particle phase space,
only the $2\to 4$ part is affected by the following modifications. 
We consider three types of uncertainties
(see \citere{De00} for more details):
\begin{itemize}
\item Different on-shell projections: \\
For the DPA
one has to specify a projection of the physical momenta to a set of 
momenta for on-shell W-pair production and decay%
\footnote{This option only illustrates the effect of different on-shell
projections in the four-particle phase space.
%, since in {\sc RacoonWW} the DPA is only applied to the virtual
%corrections.
If real photonic
corrections were treated in DPA the impact of different projections could
be larger.}, 
in order to define a DPA. 
This can be done in an obvious way by fixing the direction
of one of the \PW~bosons and of one of the final-state fermions
originating from either \PW~boson in the CM frame of the incoming
$\Pep\Pem$ pair. The default in {\sc RacoonWW}
is to fix the directions of the 
momenta of the fermions (not of the anti-fermions) resulting from the 
$\PWp$ and $\PWm$ decays (``def''). A different projection is obtained 
by fixing the direction of the anti-fermion from the $\PWp$
decay (``proj'') instead of the fermion direction.
\item Treatment of soft photons: \\
As explained above, the matching of IR and mass singularities
between virtual and real corrections requires a redistribution of the
singular parts. This redistribution fixes only
the universal, singular terms, while the redistribution of non-singular
terms are mere convention. Owing to the asymmetric treatment of the 
corrections (virtual in DPA, real from full matrix elements), different
redistributions of non-singular contributions change the result by terms
of the order $(\al/\pi)(\GW/\MW)$, 
\ie these redistributions are equivalent within the accuracy of the DPA.
In {\sc RacoonWW} two different schemes for this redistribution are
implemented. As default, the endpoint contributions of the
subtraction functions, as defined in \citere{subtract}, are calculated
in DPA and added to the virtual photon contribution. In the other
scheme, the universal IR-sensitive part is extracted from the
virtual photon contribution \`a la YFS~\cite{ye61} and added to the
real photon contribution. The resulting soft+virtual part of the
photonic correction is, thus, treated off shell (``eik'').
The difference between the two described treatments is that certain
terms of the form $(\alpha/\pi)\times\pi^2\times {\cal O}(1)$ are
either multiplied with the DPA (``def'') or with the full off-shell
Born cross sections (``eik'').
\item On-shell vs.\ off-shell Coulomb singularity: \\
The Coulomb singularity is (up to higher orders) 
fully contained in the virtual
${\cal O}(\alpha)$ correction in DPA. Performing the DPA
to the full virtual correction leads to the on-shell Coulomb
singularity, which is a simple factor of $\alpha\pi/(2\beta)$,
where $\beta$ is the velocity of an on-shell W~boson.
However, since the Coulomb singularity is an important
correction in the LEP2 energy range and is also known beyond DPA
\cite{coul},
{\sc RacoonWW} includes this extra off-shell Coulomb correction 
factor as default.
This replacement of the Coulomb singularity is performed by adding and
subtracting the corresponding contributions in the virtual non-factorizable 
corrections, as described in \citere{de98}.
Switching the extra off-shell Coulomb terms off (``Coul''), yields an
effect of the order of the uncertainty of the DPA.
\end{itemize}

In the following table and figures the total cross section
and two distributions for
$\Pep\Pem\to\Pu\bar\Pd\mu^-\bar\nu_\mu(\gamma)$
have been compared for the different versions 
of the DPA defined above.
The results have been obtained using a photon-recombination procedure
that is similar to the one decribed in the previous section. The precise
definition can be found in \citere{De00}.

The results for the total cross section are shown in
\refta{tab:sigma_DPA_unc}.
\begin{table}
\centerline{
\begin{tabular}{|c||c|c|c|}
\hline
$\sqrt{s}/\GeV$ & $\delta_{\mathrm{proj}}/\%$ & $\delta_{\mathrm{eik}}/\%$ 
& $\delta_{\mathrm{Coul}}/\%$ \\
\hline \hline
172 & $-0.03$ & $-0.09$ & 0.79 \\
\hline
200 & $-0.03$ & $-0.01$ & 0.13 \\
\hline
500 & $-0.01$ & $\phantom{-}0.08$  & 0.01 \\
\hline
\end{tabular}
}
\vspace*{1em}
\caption{Intrinsic DPA ambiguities $\delta=\sigma/\sigma_{\mathrm{def}}-1$
of the {\sc RacoonWW} predictions for the total cross section of
$\Pep\Pem\to\Pu\bar\Pd\mu^-\bar\nu_\mu(\gamma)$
(based on the results of \citere{De00})}
\label{tab:sigma_DPA_unc}
\end{table}
We find relative differences at the level of $0.1\%$.
As expected, the prediction that is based on the on-shell Coulomb
correction is somewhat higher than the exact off-shell treatment, since
off-shell effects screen the positive Coulomb singularity.
Note that for the low CM energy of $172\GeV$ the difference between
on-shell and off-shell Coulomb singularity, which is quite large
(0.79\%), cannot be viewed as a measure of the theoretical uncertainty,
since the on-shell Coulomb singularity is not adequate near threshold.

\begin{figure}[p]
{\centerline{
\setlength{\unitlength}{1cm}
\begin{picture}(6.5,6.8)
\put(-5,-15.0){\includegraphics{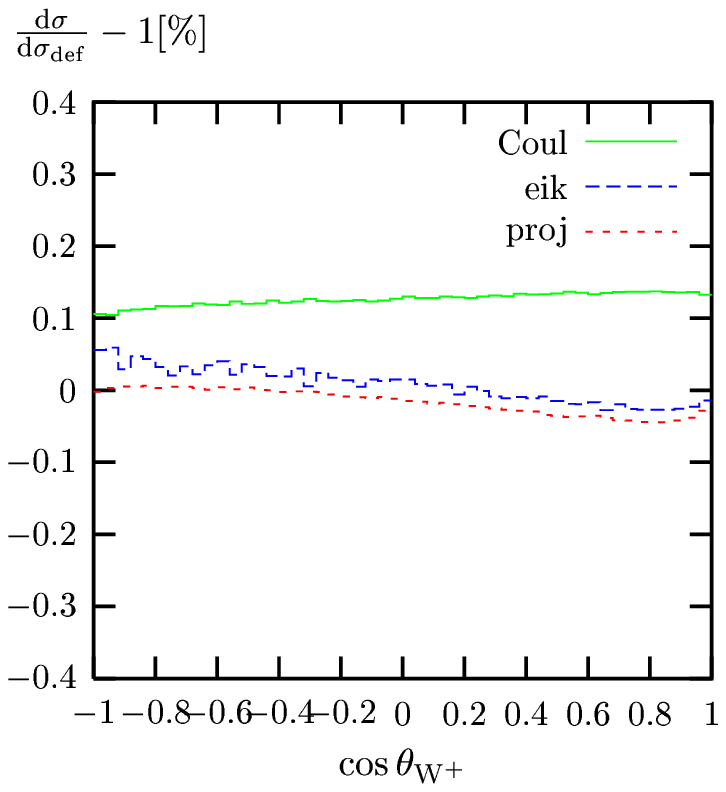}}
\end{picture} }}
\caption{Theoretical uncertainty within the DPA for distributions in the
W-production angle for
$\Pep\Pem\to\Pu\bar\Pd\mu^-\bar\nu_\mu(\gamma)$ at $\sqrt{s}=200\GeV$
(based on the results \citere{De00})}
\label{fi:tu_thwp}
\efi
\begin{figure}[p]
{\centerline{
\setlength{\unitlength}{1cm}
\begin{picture}(6.5,6.8)
\put(-5,-15.0){\includegraphics{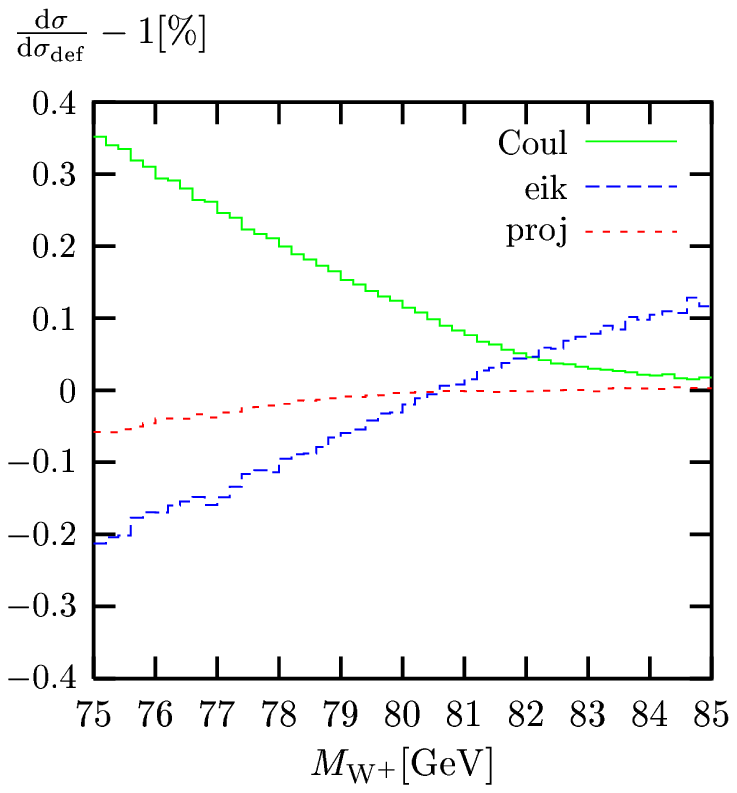}}
\end{picture} }}
\caption{Theoretical uncertainty within the DPA for distributions in
  the $\Pu\bar\Pd$ invariant mass for
  $\Pep\Pem\to\Pu\bar\Pd\mu^-\bar\nu_\mu(\gamma)$ at
  $\sqrt{s}=200\GeV$
  (based on the results \citere{De00})}
\label{fi:tu_mm}
\efi
In \reffis{fi:tu_thwp} and \ref{fi:tu_mm} we show the
differences of the ``proj'', ``eik'', and ``Coul'' modifications to
the default version of the DPA for the distributions 
in the W-production angle $\theta_{\PW^+}$ and in the 
$\Pu\bar\Pd$ invariant-mass $M_{\Pu\bar\Pd}$, 
both at $\sqrt{s}=200\GeV$.  For the $\theta_{\PW^+}$ distribution 
the relative differences are of the order of $0.1$--$0.2\%$ for all
angles, which is of the expected order for the intrinsic DPA uncertainty. 
For the $M_{\Pu\bar\Pd}$ distribution
the DPA uncertainties are at the level of $0.1$--$0.3\%$ within a window
of $2\Gamma_\PW$ around the W~resonance. The uncertainties
grow with the distance from the resonance point, as expected, since the
DPA runs out of its range of validity away from the resonance region.


\vspace*{-.5em}
\begin{thebibliography}{99}
\frenchspacing
\newcommand{\ap}[3]{{\sl Ann.\ Phys.} {\bf #1} (#2) #3}
\newcommand{\app}[3]{{\sl Acta.\ Phys.\ Pol.} {\bf #1} (#2) #3}
\newcommand{\zp}[3]{{\sl Z.\ Phys.} {\bf #1} (#2) #3}
\newcommand{\np}[3]{{\sl Nucl.\ Phys.} {\bf #1} (#2) #3}
\newcommand{\pl}[3]{{\sl Phys.\ Lett.} {\bf #1} (#2) #3}
\newcommand{\pr}[3]{{\sl Phys.\ Rev.} {\bf #1} (#2) #3}
\newcommand{\prl}[3]{{\sl Phys.\ Rev.\ Lett.} {\bf #1} (#2) #3}
\newcommand{\epjdir}[3]{{\sl EPJdirect} {\bf #1} (#2) #3}
\newcommand{\fp}[3]{{\sl Fortschr.\ Phys.} {\bf #1} (#2) #3}
\newcommand{\jp}[3]{{\sl J.\ Phys.} {\bf #1} (#2) #3}
\newcommand{\cpc}[3]{{\sl Comput.\ Phys.\ Commun.} {\bf #1} (#2) #3}
\newcommand{\ijmp}[3]{{\sl Int.\ J.\ Mod.\ Phys.} {\bf #1} (#2) #3}
\newcommand{\nim}[3]{{\sl Nucl.\ Instr.\ Meth.} {\bf #1} (#2) #3}
\newcommand{\nc}[3]{{\sl Nuovo Cimento} {\bf #1} (#2) #3}
\newcommand{\vj}[4]{{\sl #1} {\bf #2} (#3) #4}

\bibitem{lep2res}
G.\ Quast, these proceedings.

\bibitem{lep2repWcs}
W.\ Beenakker et al., 
in {\sl Physics at LEP2} (Report CERN 96-01, Geneva, 1996),
G.\ Altarelli, T.\ Sj\"o\-strand and F.\ Zwirner (eds.),
Vol.\ 1, p.\ 79, hep-ph/9602351.

\bibitem{lep2mcws}
{\it Four-Fermion Working Group Report} of the {\it LEP2 Monte Carlo
Workshop}, CERN, 1999/2000, in preparation.

\bibitem{st91}
R.G.\ Stuart, \pl{B262}{1991}{113};\\
A.\ Aeppli, G.J.\ van\ Oldenborgh and D.\ Wyler, \np{B428}{1994}{126}.

\bibitem{yfsww}
S.\ Jadach et al., \pl{B417}{1998}{326}; hep-ph/9907436.

\bibitem{Be98}
W. Beenakker,  F.A. Berends and A.P. Chapovsky, \np{B548}{1999}{3}.

\bibitem{ku99} 
Y. Kurihara, M. Kuroda and D. Schildknecht,
\np{B565}{2000}{49}.

\bibitem{racoonww_lep2res}
A. Denner, S. Dittmaier, M. Roth and D. Wackeroth,
\pl{B475}{2000}{127} and
\epjdir{C4}{2000}{1} (hep-ph/9912447).

\bibitem{De00}
A. Denner, S. Dittmaier, M. Roth and D. Wackeroth, BI-TP 2000/06,
hep-ph/0006307.

\bibitem{wwprod}
%M. B\"ohm, A. Denner, T. Sack, W. Beenakker, F. Berends and H. Kuijf,
M. B\"ohm et al.,
\np{B304}{1988}{463}; \\
J. Fleischer, F. Jegerlehner and M. Zralek,
\zp{C42}{1989}{409}.

\bibitem{wdecay}
A. Denner and T. Sack,
\zp{C46}{1990}{653}; \\
D.Y. Bardin, S. Riemann and T. Riemann,
\zp{C32}{1986}{121}; \\
F. Jegerlehner,
\zp{C32}{1986}{425}.

\bibitem{be97}
W. Beenakker, F.A. Berends and A.P. Chapovsky,
\pl{B411}{1997}{203} and \np{B508}{1997}{17}.

\bibitem{de98}
A.\ Denner, S.\ Dittmaier and M.\ Roth, 
\np{B519}{1998}{39} and \pl{B429}{1998}{145}.

\bibitem{ee4fa}
A. Denner, S. Dittmaier, M. Roth and D. Wackeroth, 
\np{B560}{1999}{33}.

\bibitem{subtract}
S.\ Dittmaier, \np{B565}{2000}{69};\\
M. Roth, dissertation ETH Z\"urich No.\ 13363, 1999.

\bibitem{LEPEWWG}
%Homepage of the LEP Electroweak Working Group, \\
http://lepewwg.web.cern.ch/LEPEWWG/.

\bibitem{gentle}
D.\ Bardin, M.\ Bilenky, A.\ Olchevski and T.\ Riemann,
\pl{B308}{1993}{403};\\
D.\ Bardin et al., \cpc{104}{1997}{161}.

\bibitem{ye61}
D.R. Yennie, S.C. Frautschi and H. Suura, \ap{13}{1961}{379}.

\bibitem{coul}
V.S.\ Fadin, V.A.\ Khoze and A.D.\ Martin, \pl{B311}{93}{311};\\
D.\ Bardin, W.\ Beenakker and A.\ Denner, \pl{B317}{93}{213};\\
V.S.\ Fadin et al., \pr{D52}{95}{1377}.

\end{thebibliography}
\end{document}